\documentstyle[12pt]{article}

\newcommand{\cl}{\centerline}

\newcommand\beq{\begin{equation}}
\newcommand\eeq{\end{equation}}
\newcommand\bea{\begin{eqnarray}}
\newcommand\eea{\end{eqnarray}}

\begin{document}

\begin{titlepage}
\setlength{\textwidth}{5.0in}
\setlength{\textheight}{7.5in}
\setlength{\parskip}{0.0in}
\setlength{\baselineskip}{18.2pt}
\hfill SOGANG-HEP 249/98
\begin{center}
\cl{\Large{{\bf Consistent Dirac Quantization of}}}\par
\cl{\Large{{\bf SU(2) Skyrmion equivalent to the BFT Scheme}}}\par
\vskip 0.6cm
\begin{center}
{Soon-Tae Hong$^{1}$, Yong-Wan Kim$^{1,2}$ and Young-Jai Park$^{1}$}\par
\end{center}
\vskip 0.3cm
\begin{center}
{${}^{1}$Department of Physics and Basic Science Research Institute,}\par
{Sogang University, C.P.O. Box 1142, Seoul 100-611, Korea}\par
\vskip 0.2cm
{${}^{2}$Department of Physics Education,}\par
{Seoul National University, Seoul 151-742, Korea}
\end{center}
\vskip 0.3cm
\cl{\today}
\vskip 0.5cm

\begin{center}
{\bf ABSTRACT}
\end{center}
\begin{quotation}
In the framework of Dirac quantization, the SU(2) Skyrmion is canonically 
quantized to yield the modified predictions of the static properties of 
baryons.  We show that the energy spectrum of this Skyrmion obtained by 
the Dirac method with the suggestion of generalized momenta is consistent with 
the result of the Batalin-Fradkin-Tyutin formalism.
\vfill
\noindent
\end{quotation}
\end{center}
\end{titlepage}

It is well known that baryons can be obtained from topological solutions,
known as SU(2) Skyrmions, since the homotopy group $\Pi_{3}(SU(2))=Z$ admits
fermions \cite{ad,sk,hsk}. Using the collective coordinates of the isospin
rotation of the Skyrmion, Adkins et al. \cite{ad} have performed
semiclassical quantization having the static properties of baryons within
30$\%$ of the corresponding experimental data. Also the chiral bag model, 
which is a hybrid of two different models, the MIT bag model at infinite bag 
radius on one hand, and the SU(3) Skyrmion model at a vanishing radius on the 
other hand, has enjoyed considerable success in predicting the strange 
form factors of baryons \cite{sff} 
to confirm the recent experimental result of 
the SAMPLE Collaboration \cite{sam}.

On the other hand, in order to quantize the physical systems subjective to
the constraints, the Dirac quantization scheme \cite{di} has been used
widely. First of all, string theory is known to be restricted to obey the
Virasoro conditions, and thus it is quantized by the Dirac method \cite{gr}.
Also, in the 2+1 dimensional O(3) $\sigma$ model, Bowick {\it et al.} 
\cite{bo} have used the Dirac scheme to obtain the fractional spin.

However, whenever we adopt the Dirac method, we frequently meet the problem
of the operator ordering ambiguity. In order to avoid this problem, Batalin,
Fradkin, and Tyutin (BFT) developed a method \cite{BFT} which converts the
second-class constraints into first-class ones by introducing auxiliary
fields. Recently, this BFT formalism has been applied to several interesting
models \cite{BFT1}. Very recently, the SU(2) Skyrme model has been studied
in the context of the Abelian and non-Abelian BFT formalism \cite{sk2,skn}.
But, there exists some inconsistency on the constraint structure.

In this paper, we will canonically quantize the SU(2) Skyrme model by using
the desired Dirac quantization method, which is consistent with the BFT one.
Firstly, the Dirac bracket scheme will be discussed in the framework of the
SU(2) Skyrmions to quantize the baryons. The adjustable parameter will be
introduced to define the generalized momenta without any loss of generality.
Next, we will apply the proper BFT method to the Skyrmion to obtain the
energy spectrum of the baryons by including the Weyl ordering correction.
Finally, we will show that by fixing this free parameter the baryon
energy eigenvalues obtained by the Dirac method are consistent with the
result of the BFT formalism  and modify the predictions of the baryon static 
properties.

Now we start with the Skyrmion Lagrangian of the form 
\begin{equation}
L=\int{\rm d}r^{3}\left[\frac{f_{\pi}^{2}}{4}{\rm tr}(\partial_{\mu}U^{%
\dag}\partial^{\mu}U)+\frac{1} {32e^{2}}{\rm tr}[U^{\dag}\partial_{\mu}U,U^{%
\dag}\partial_{\nu}U]^{2}\right],
\end{equation}
where $f_{\pi}$ is the pion decay constant, $e$ is a dimensionless
parameter, and $U$ is an SU(2) matrix satisfying the boundary condition $%
\lim_{r \rightarrow \infty} U=I$, so that the pion field vanishes as $r$ goes
to infinity. For the minimum energy of the Skyrmion, one can take the
hedgehog ansatz $U_{0}(\vec{x})=e^{i\tau_{a}\hat{x}_{a}f(r)}$, where the $%
\tau_{a}$ are Pauli matrices, $\hat{x}=\vec{x}/r$ and for the unit winding
number $\lim_{r \rightarrow \infty} f(r)=0$ and $f(0)=\pi$. On the other
hand, since the hedgehog ansatz has maximal or spherical symmetry, it is
easily seen that spin plus isospin equals zero, so that isospin
transformations and spatial rotations are related to each other.

Furthermore, in the Skyrmion model, spin and isospin states can be treated
by collective coordinates $a^{\mu}=(a^{0},\vec{a})$ $(\mu=0,1,2,3)$
corresponding to the spin and isospin rotations $A(t) = a^{0}+i\vec{a}\cdot
\vec{\tau}$. With the hedgehog ansatz and the collective rotation $A(t)\in$ 
SU(2), the chiral field can be given by $U(\vec{x},t)=A(t)U_{0}(\vec{x})
A^{\dagger}(t)=e^{i\tau_{a}R_{ab}\hat{x}_{b}f(r)}$ where $R_{ab}=\frac{1}{2}
{\rm tr} (\tau_{a}A\tau_{b}A^{\dagger})$.

The Skyrmion Lagrangian is then given by 
\footnote{Here one can easily check that the Skyrmion Lagrangian 
can be rewritten as 
$L=-E+2{\cal I}\vec{\alpha}^{2}$ by defining the new variables 
$\alpha^{k}=a^{0}\dot{a}^{k}-\dot{a}^{0}a^{k}+\epsilon^{kpq}a^{p}\dot{a}^{q}$}.
\begin{equation}
L=-E+2{\cal I}\dot{a}^{\mu}\dot{a}^{\mu},
\label{lag}
\end{equation}
where the soliton energy and the moment of inertia are given by 
\begin{eqnarray}
E&=&4\pi\int_{0}^{\infty}{\rm d}r r^{2}
\left[ \frac{f_\pi^2}{2}
     [ (\frac{{\rm d}f}{{\rm d}r})^{2}+2\frac{\sin^{2}f}{r^{2}}] 
    +\frac{1}{2e^2} \frac{\sin^{2}f}{r^2} 
[2(\frac{{\rm d}f}{{\rm d}r})^{2}+\frac{\sin^{2}f}{r^{2}}]\right],
\nonumber\\
{\cal I}&=&\frac{8\pi}{3}\int_{0}^{\pi}{\rm d}r r^{2}\sin^{2}f
\left[f_{\pi}^{2}+ \frac{1}{e^{2}}
((\frac{{\rm d}f}{{\rm d}r})^{2}+\frac{\sin^{2}f}{r^{2}})\right].
\label{eni}
\end{eqnarray}
Introducing the canonical momenta $\pi^{\mu}=4{\cal I}\dot{a}^{\mu}$ conjugate 
to the collective coordinates $a^{\mu}$ one can then obtain the canonical 
Hamiltonian 
\begin{equation}
H=E+\frac{1}{8{\cal I}}\pi^{\mu}\pi^{\mu}
\label{hamil}
\end{equation}
and the spin and isospin operators 
\begin{eqnarray}
J^{i}&=&\frac{1}{2}(a^{0}\pi^{i}-a^{i}\pi^{0}-\epsilon_{ijk}a^{j}\pi^{k}), 
\nonumber \\
I^{i}&=&\frac{1}{2}(a^{i}\pi^{0}-a^{0}\pi^{i}-\epsilon_{ijk}a^{j}\pi^{k}).
\label{spin}
\end{eqnarray}

On the other hand, we have the following second-class constraints:\footnote{%
Here one notes that, due to the commutator $\{\pi^{\mu},\Omega_{1}\}
=-2a^{\mu}$, one can obtain the algebraic relation $\{\Omega_1,H\}={\frac{1}{%
2{\cal {I}}}}\Omega_2$.} 
\begin{eqnarray}
\Omega_{1} &=& a^{\mu}a^{\mu}-1\approx 0,  \nonumber \\
\Omega_{2} &=& a^{\mu}\pi^{\mu}\approx 0,  \label{omega2}
\end{eqnarray}
to yield the Poisson algebra 
\begin{equation}
\Delta_{k k^{\prime}}=\{\Omega_{k},\Omega_{k^{\prime}}\} = 2\epsilon^{k
k^{\prime}}a^{\mu}a^{\mu}  
\label{delta}
\end{equation}
with $\epsilon^{12}=-\epsilon^{21}=1$.  Using the Dirac brackets \cite{di} 
defined by 
$$
\{A,B\}_{D}=\{A,B\}-\{A,\Omega_{k}\}\Delta^{k k^{\prime}}
\{\Omega_{k^{\prime}},B\}
$$ 
with $\Delta^{k k^{\prime}}$ being the inverse of $\Delta_{k k^{\prime}}$ and 
performing the canonical quantization scheme $\{A,B\}_{D}\rightarrow 
\frac{1}{i}[A_{op},B_{op}]$ one can obtain the operator commutators 
\begin{eqnarray}
{[a^{\mu},a^{\nu}]}&=&0, \nonumber \\
{[a^{\mu},\pi^{\nu}]}&=&i(\delta^{\mu \nu}-\frac{a^{\mu}a^{\nu}}{a^{\sigma}
                         a^{\sigma}}),  \nonumber \\
{[\pi^{\mu},\pi^{\nu}]}&=&\frac{i}{a^{\sigma}a^{\sigma}}
                         (a^{\nu}\pi^{\mu}-a^{\mu}\pi^{\nu})  
\label{aaop}
\end{eqnarray}
with $\pi^{\mu}=-i(\delta^{\mu\nu}-\frac{a^{\mu}a^{\nu}}{a^{\sigma}a^{\sigma}})
\partial_{\nu}$, and the closed current algebra 
\begin{eqnarray}
{[M^{\mu},M^{\nu}]}&=&\epsilon^{\mu \nu \sigma}M^{\sigma},  \nonumber \\
{[M^{\mu},N^{\nu}]}&=&\epsilon^{\mu \nu \sigma}N^{\sigma},  \nonumber \\
{[N^{\mu},N^{\nu}]}&=&0
\nonumber
\end{eqnarray}
with $M^{\mu}=i\epsilon^{\mu \nu \sigma}\pi^{\nu}a^{\sigma}$, $%
N^{\mu}=ia^{\mu}$.

Now we observe that without any loss of generality the generalized momenta $%
\Pi^{\mu}$ fulfilling the structure of the commutators (\ref{aaop}) are of the 
form \footnote{In Ref. \cite{fujii} the authors did not include the last term 
so that one cannot clarify the relations between the BFT scheme and the Dirac 
bracket one. Also one can easily see that $\Pi^{\mu}$ are not the canonical 
momenta conjugate to the collective coordinates $a^{\mu}$ any more since 
$\Pi^{\mu}$ depend on $a^{\mu}$, as expected.} 
\begin{equation}
\Pi^{\mu}=-i(\delta^{\mu\nu}-\frac{a^{\mu}a^{\nu}}{a^{\sigma}a^{\sigma}})
\partial_{\nu}-\frac{ic a^{\mu}}{a^{\sigma}a^{\sigma}}
\label{cterm}
\end{equation}
with an arbitrary parameter $c$ to be fixed later. It does not also change
the spin and isospin operators (\ref{spin}).

On the other hand, the energy spectrum of the baryons in the SU(2) Skyrmion
can be obtained in the Weyl ordering scheme \cite{weyl} where the
Hamiltonian (\ref{hamil}) is modified into the symmetric form 
\begin{equation}
H_{N}=E+\frac{1}{8{\cal I}}\Pi^{\mu}_{N}\Pi^{\mu}_{N},  \label{hamil2}
\end{equation}
where 
\begin{equation}
\Pi^{\mu}_{N}=-\frac{i}{2}
\left[(\delta^{\mu\nu}-\frac{a^{\mu}a^{\nu}}{a^{\sigma}
a^{\sigma}})\partial_{\nu}
+\partial_{\nu}(\delta^{\mu\nu}-\frac{a^{\mu}a^{\nu}}{a^{\sigma}a^{\sigma}})
+\frac{2c a^{\mu}}{a^{\sigma}a^{\sigma}}\right].
\end{equation}
After some algebra, one can obtain the Weyl ordered $\Pi^{\mu}_{N}
\Pi^{\mu}_{N}$ as follows:\footnote{
Here the first three terms are nothing but the three-sphere Laplacian \cite
{corr} given in terms of the collective coordinates and their derivatives to
yield the eigenvalues $l(l+2)$.}
$$
\Pi^{\mu}_{N}\Pi^{\mu}_{N}=-\partial_{\mu}\partial_{\mu}
+\frac{3a^{\mu}}{a^{\sigma}a^{\sigma}}\partial_{\mu}
+\frac{a^{\mu}a^{\nu}}{a^{\sigma}a^{\sigma}}\partial_{\mu}\partial_{\nu}
+\frac{1}{a^{\sigma}a^{\sigma}}(\frac{9}{4}-c^{2})
$$
to yield the modified quantum energy spectrum of 
the baryons\footnote{Due to the missing factor 
$a^{\sigma}a^{\sigma}$ in the denominators 
in Eq. (\ref{aaop}) which is ignored in Refs. \cite{fujii,corr}, apart from 
$-c^{2}$ originated from the additional $c$-term in Eq. (\ref{cterm}) 
we obtain 
the Weyl ordering correction $\frac{9}{4}$, different from the value 
$\frac{5}{4}$ given in Ref. \cite{corr}.}    
\begin{equation}
\langle H_{N}\rangle=E+\frac{1}{8{\cal I}}
\left[l(l+2)+\frac{9}{4}-c^{2}\right].
\label{hwc}
\end{equation}

Next, following the Abelian BFT formalism \cite{BFT,BFT1,sk2} which 
systematically converts the second-class constraints into first-class ones,
we introduce two auxiliary fields $\Phi^{i}$ corresponding to $\Omega_{i}$
with the Poisson brackets 
\begin{equation}
\{\Phi^{i}, \Phi^{j}\}=\omega^{ij}.
\end{equation}
The first-class constraints $\tilde{\Omega}_{i}$ are then constructed as a 
power series of the auxiliary fields: 
\begin{equation}
\tilde{\Omega}_{i}=\sum_{n=0}^{\infty}\Omega_{i}^{(n)},~~~~
\Omega_{i}^{(0)}=\Omega_{i}  \label{tilin}
\end{equation}
where $\Omega_{i}^{(n)}$ are polynomials in the auxiliary fields $\Phi^{j}$
of degree $n$, to be determined by the requirement that the first-class
constraints $\tilde{\Omega}_{i}$ satisfy an Abelian algebra as follows:
\begin{equation}
\{\tilde{\Omega}_{i},\tilde{\Omega}_{j}\}=0.  \label{cijk}
\end{equation}
Since $\Omega_{i}^{(1)}$ are linear in the auxiliary fields, one can make
the ansatz 
\begin{equation}
\Omega_{i}^{(1)}=X_{ij}\Phi^{j}.  \label{xijphi}
\end{equation}
Substituting Eq. (\ref{xijphi}) into Eq. (\ref{cijk}) leads to the following
relation:
\begin{equation}
\Delta_{ij}+X_{ik}\omega^{kl}X_{jl}=0,  \label{delx}
\end{equation}
which, for the standard choice \cite{BFT,BFT1,sk2} of $\omega^{ij}=
\epsilon^{ij}$, has a solution 
\begin{equation}
X_{ij}=\left( 
\begin{array}{cc}
2 & 0 \\ 
0 & -a^{\mu}a^{\mu}
\end{array}
\right).  \label{xij}
\end{equation}
Substituting Eq. (\ref{xij}) into Eqs. (\ref{tilin}) and (\ref{xijphi}) and 
iterating this procedure, one can obtain the first-class constraints 
\begin{eqnarray}
\tilde{\Omega}_{1}&=&\Omega_{1}+2\Phi^{1},  \nonumber \\
\tilde{\Omega}_{2}&=&\Omega_{2}-a^{\mu}a^{\mu}\Phi^{2},  \label{1stconst}
\end{eqnarray}
which yield the strongly involutive first-class constraint algebra (\ref
{cijk}). On the other hand, the corresponding first-class Hamiltonian is
given by 
\begin{equation}
\tilde{H}= E +\frac{1}{8{\cal I}}(\pi^{\mu}-a^{\mu}\Phi^{2})
(\pi^{\mu}-a^{\mu}\Phi^{2})\frac{a^{\nu}a^{\nu}}{a^{\nu}a^{\nu}+2\Phi^{1}},
\label{hct}
\end{equation}
which is also strongly involutive with the first-class constraints 
$$
\{\tilde{\Omega}_{i},\tilde{H}\}=0.
$$
Here one notes that, with the Hamiltonian (\ref{hct}), one cannot naturally
generate the first-class Gauss' law constraint from the time evolution of
the primary constraint $\tilde{\Omega}_{1}$. Now, by introducing an
additional term proportional to the first-class constraints $\tilde{\Omega}%
_{2}$ into $\tilde{H}$, we obtain an equivalent first-class Hamiltonian 
\begin{equation}
\tilde{H}^{\prime}=\tilde{H}+\frac{1}{4{\cal I}}\Phi^{2} \tilde{\Omega}_{2},
\label{hctp}
\end{equation}
which naturally generates the Gauss' law constraint 
\begin{eqnarray}
\{\tilde{\Omega}_{1},\tilde{H}^{\prime}\}&=&\frac{1}{2{\cal I}} 
\tilde{\Omega}_{2},  \nonumber \\
\{\tilde{\Omega}_{2},\tilde{H}^{\prime}\}&=&0.
\end{eqnarray}
Here one notes that $\tilde{H}$ and $\tilde{H}^{\prime}$ act on physical 
states in the same way since such states are annihilated by the first-class 
constraints. Similarly, the equations of motion for observables are also 
unaffected by this difference. Furthermore, if we take the limit 
$\Phi^{i}\rightarrow 0$, then our first-class system exactly returns to the 
original second-class one.

Now, using the first-class constraints in the Hamiltonian (\ref{hctp}), one
can obtain a Hamiltonian of the form\cite{sk2} 
\begin{equation}
\tilde{H}^{\prime}=E+\frac{1}{8{\cal I}}(a^{\mu}a^{\mu}\pi^{\nu}\pi^{\nu}
-a^{\mu}\pi^{\mu}a^{\nu}\pi^{\nu}).  \label{htilde}
\end{equation}
Following the symmetrization procedure, the first-class Hamiltonian yields
the energy spectrum with the Weyl ordering correction  
\begin{equation}
\langle\tilde{H}^{\prime}_{N}\rangle=E+\frac{1}{8{\cal I}}[l(l+2)+1].
\label{nht}
\end{equation}
Then, in order for the Dirac bracket scheme to be consistent with the BFT
one, the adjustable parameter $c$ in Eq. (\ref{hwc}) should be fixed with the
values 
\begin{equation}
c=\pm\frac{\sqrt{5}}{2}.
\end{equation}
Here one notes that these values for the parameter $c$ relate the Dirac 
bracket scheme with the BFT one to yield the desired quantization in the 
SU(2) Skyrmion model so that one can achieve the unification of these two 
formalisms.

Next, using the Weyl ordering corrected energy spectrum (\ref{nht}), we easily 
obtain the hyperfine structure of the nucleon and $\Delta$ hyperon masses to
yield the soliton energy and the moment of inertia 
\begin{eqnarray}
E&=&\frac{1}{3}(4M_{N}-M_{\Delta})\nonumber\\
{\cal I}&=&\frac{3}{2}(M_{\Delta}-M_{N})^{-1}.
\label{masses}
\end{eqnarray}
Substituting the experimental values $M_{N}=939$ MeV and $N_{\Delta}=1232$ MeV 
into Eq. (\ref{masses}) and using expressions (\ref{eni}), one can predict 
the pion decay constant $f_{\pi}$ and the Skyrmion parameter $e$ as follows:
$$
f_{\pi}=63.2~{\rm MeV},~~~e=5.48.
$$  
With these fixed values of $f_{\pi}$ and $e$, one can then proceed to yield the 
predictions for the other static properties of the baryons.  The isoscalar and 
isovector mean-square (magnetic) charge radii and the baryon and transition 
magnetic moments are contained in Table 1, together with the experimental 
data and the standard Skyrmion predictions \cite{ad,hsk,liu}.\footnote{For the 
$\Delta$ magnetic moments, we use the experimental data of Nefkens 
{\it et al}. \cite{nef}.} 
It is remarkable that the effects of Weyl ordering correction 
in the baryon energy spectrum are propagated through the model parameters 
$f_{\pi}$ and $e$ to modify the predictions of the baryon static properties. 

It seems appropriate to comment on the ``non-Abelian" BFT scheme of this
Skyrme model, although this scheme gives the same baryon energy eigenvalues 
\cite{skn}. This non-Abelian scheme is mainly based on the introduction
of auxiliary fields satisfying 
\begin{eqnarray}  
\{\tilde\Omega_i, \tilde\Omega_j\}&=&C^k_{ij}\tilde\Omega_k,\nonumber\\
\{\tilde\Omega_i, \tilde{H}\}&=&B^j_i\tilde\Omega_j,\label{nac2}
\end{eqnarray}
where $\tilde\Omega_i$ and $\tilde{H}$ can be constructed as a power series of
auxiliary fields as before. Then, besides $\omega^{ij}$ and $X_{ij}$ to be
chosen, one should find the coefficients $C^k_{ij}$ further, which solve $%
C^k_{ij}\Omega_k=\Delta_{ij}+X_{ik}\omega^{kl}X_{jl}$ at the zeroth order of
Eq. (\ref{nac2}). Among many possible values, if one chooses $C^1_{12}=2$, $%
\omega^{12}=-\omega^{21}=1$, $X_{11}=-X_{22}=1$ with the other vanishing
components as in Ref. \cite{skn}, one would have the first-class constraints
having a nonlinear term of auxiliary fields as 
\begin{eqnarray}
\tilde\Omega_1&=&\Omega_1+\Phi^1  \nonumber \\
\tilde\Omega_2&=&\Omega_2-\Phi^2+\Phi^1\Phi^2
\end{eqnarray}
satisfying the constraint algebra 
\begin{eqnarray}
&&\{\tilde\Omega_1,\tilde\Omega_1\}=\{\tilde\Omega_2,\tilde\Omega_2\}=0,\\
&& \{\tilde\Omega_1,\tilde\Omega_2\}=2\tilde\Omega_1. \label{otwo}
\end{eqnarray}
Moreover, using the corresponding first-class Hamiltonian such as 
\begin{eqnarray}
\tilde{H}&=& H -\frac{1}{8{\cal I}}\pi^{\mu}\pi^{\mu}\Phi^1 +\frac{1}{2}%
(B^1_1\Omega_1-\frac{1}{2{\cal I}}\Omega_2)\Phi^2  \nonumber \\
&&+\frac{1}{2}(B^1_1+\frac{1}{2{\cal I}}\Omega_2)\Phi^1\Phi^2 +\frac{1}{8%
{\cal I}}a^\mu a^\mu(1-\Phi^1) \Phi^2\Phi^2,  \label{ht}
\end{eqnarray}
we obtain 
\begin{eqnarray}  \label{incon2}
\{\tilde\Omega_1, \tilde{H}\}&=&B^1_1\tilde\Omega_1,  \nonumber \\
\{\tilde\Omega_2, \tilde{H}\}&=&0,
\end{eqnarray}
where $B^1_1$ remains undetermined in general. This non-Abelian scheme
seems to work, {\it i.e.,} the first-class Hamiltonian (\ref{ht}) has simple
finite sums for this nonlinear theory, compared with the previous one Eq. (\ref
{hct}), and thus it would be an adequate approach to studying such a
nonlinear theory rather than the Abelian version of BFT.

However, there still exists some inconsistency in the algebraic relations, 
which
should be resolved, even though the Hamiltonian (\ref{ht}) yields the same 
energy eigenvalues (\ref{nht}) as in the Abelian case.  In particular, Eq. 
(\ref{otwo}) in the first-class constraint algebra is not consistent in the 
limit of the auxiliary fields $\Phi^i\rightarrow 0$, {\it i.e.,} it does not 
recover the original second-class structure such as the Poisson algebra 
(\ref{delta}). This kind of situation happens again when one considers Eq. 
(\ref{incon2}) obtained from the non-Abelian BFT scheme. Moreover, it does not 
generate the Gauss' law constraint naturally.

In summary, we have clarified the relation between the Dirac bracket scheme
and the BFT one, which has been obscure and unsettled, in the framework of the
SU(2) Skyrmion model. In this approach we have introduced the generalized
momentum operators with the free parameter, which is fixed to yield the
consistency between these two formalisms.  We have shown that one could 
see the effects of the Weyl ordering correction in the baryon energy spectrum 
propagated through the model parameters $f_{\pi}$ and $e$ in the predictions 
of the baryon static properties.  Also, in the Abelian BFT scheme, we have 
obtained the Gauss's law constraint which was not attainable in the 
non-Abelian BFT one.  Finally, through further investigation, the SU(3) 
extension \cite{su3} of this analysis will be studied.

\vskip 1.0cm S.T.H would like to thank Professor G.E. Brown for 
helpful discussions and constant concerns. The present work was supported by
the Basic Science Research Institute Program, Korean Research Foundation,
Project No. 1998-015-D00074. Y.-W Kim was supported in part by KOSEF (1998).

\newpage
\begin{table}[t]
\caption{The static properties of baryons in the standard and Weyl ordering
corrected (WOC) Skyrmions compared with experimental data.  The 
quantities used as input parameters are indicated by $*$.}
\begin{center}
\begin{tabular}{crrr}
\hline
\hline
Quantity  &Standard  &WOC &Experiment\\
\hline
$M_{N}$ &939 {\rm MeV}$^{*}$ &939 {\rm MeV}$^{*}$ &939 {\rm MeV}\\
$M_{\Delta}$ &1232 {\rm MeV}$^{*}$ &1232 {\rm MeV}$^{*}$ &1232 {\rm MeV}\\
$f_{\pi}$ &64.5 {\rm MeV} &63.2 {\rm MeV} &93.0 {\rm MeV}\\
e &5.44 &5.48 & \\
$\langle r^{2}\rangle^{1/2}_{M,I=0}$  &0.92 {\rm fm} &0.94 {\rm fm}
                                      &0.81 {\rm fm}\\
$\langle r^{2}\rangle^{1/2}_{M,I=1}$  &$\infty$ &$\infty$ &0.80 {\rm fm}\\
$\langle r^{2}\rangle^{1/2}_{I=0}$  &0.59 {\rm fm} &0.60 {\rm fm}
                                    &0.72 {\rm fm}\\
$\langle r^{2}\rangle^{1/2}_{I=1}$  &$\infty$ &$\infty$ &0.88 {\rm fm}\\
$\mu_{p}$ &1.87 &1.89 &2.79\\
$\mu_{n}$ &$-1.31$ &$-1.32$ &$-1.91$\\
$\mu_{\Delta^{++}}$ &3.72 &3.75 &4.7$-$6.7\\
$\mu_{N\Delta}$ &2.27 &2.27 &3.29\\
$\mu_{p}-\mu_{n}$ &3.18 &3.21 &4.70\\
\hline
\hline
\end{tabular}
\end{center}
\end{table}
\end{document}